\documentclass[conference]{IEEEtran}
\IEEEoverridecommandlockouts

\usepackage{cmap}   
\usepackage[T1]{fontenc}
\usepackage{cite}
\usepackage{amsmath,amssymb,amsfonts}
\usepackage{algorithmic}
\usepackage{graphicx}
\usepackage{textcomp}
\usepackage{xcolor}

\usepackage{soul}
\usepackage{siunitx}
\usepackage[hidelinks]{hyperref}
\hypersetup{
  pdftitle={An Open-Source 8-Channel Ultrasound Transmit IP in a 130 nm RISC-V SoC},
  pdfauthor={Federico Villani, Nico Canzani, Marc-Andre Wessner, Philippe Sauter, Enrico Zelioli, Andrea Cossettini, Christoph Leitner, Luca Benini},
  pdfsubject={IEEE International Ultrasonics Symposium 2026},
  pdfkeywords={Ultrasound transmit, pulser, open-source hardware, RISC-V, system-on-chip, multi-element, 130 nm CMOS}
}
\usepackage{tikz}
\usetikzlibrary{arrows.meta, shapes.geometric}
\usepackage[nameinlink, capitalise]{cleveref}
\usepackage[nolist]{acronym}
\usepackage[flushleft]{threeparttable}
\usepackage{placeins}
\usepackage{orcidlink}
\usepackage[percent]{overpic}
\usepackage{mathtools}
\usepackage{multirow}
\usepackage{booktabs}
\usepackage{tabularx}
\usepackage{float}
\usepackage{comment}
\usepackage{enumitem}

\newcommand{\todo}[1]{}

\newcommand{\idea}[1]{}

\newcommand{\RNum}[1]{\uppercase\expandafter{\romannumeral #1\relax}}

\DeclareSIUnit\db{dB}
\DeclareSIUnit\dbm{dBm}
\DeclareSIUnit\Vpp{V\ensuremath{_{\mathrm{pp}}}}
\DeclareSIUnit\mAh{mAh}
\DeclareSIUnit\SPS{Sps}
\DeclareSIUnit\MSPS{Msps}

\begin{acronym}

    \acro{BAN}{Body Area Network}
    \acro{PAN}{Personal Area Network}
    \acro{IoB}{Internet of Bodies}
    \acro{AI}{Artificial Intelligence}
    \acro{MCU}{microcontroller}
    \acro{Nb-IoT}{Narrowband IoT}
    \acro{LoRa}{Long Range}
    \acro{UWB}{Ultra-Wideband}
    \acro{NFC}{Near Field Communication}
    \acro{VR}{Virtual Reality}
    \acro{AR}{Augmented Reality}
    \acro{EQS-HBC}{Electro-Quasistatic Human Body Communication}
    \acro{EQS}{Electro-Quasistatic}
    \acro{HBC}{Human Body Communication}
    \acro{WBAN}{Wireless Body Area Network}
    \acro{WPAN}{Wireless Personal Area Network}
    \acro{WP}{Work Package}
    \acro{CSMA}{Carrier Sense Multiple Access}
    \acro{KB}{Kilobyte}
    \acro{PMIC}{Power Management IC}

    \acro{HCI}{Human-Computer Interaction}
    \acro{HMI}{Human-Machine Interaction}
    \acro{TIA}{Transimpedance Amplifier}
    \acro{SoTA}{State-of-The-Art}
    \acro{WLAN}{Wireless Local Area Network}
    \acro{PCE}{Power Conversion Efficiency}
    \acro{ECG}{Electrocardiogram}
    \acro{BOM}{Bill of Material}
    \acro{DAQ}{Data Acquisition}
    \acro{RSSI}{Received Signal Strength Indicator}
    \acro{RSS}{Received Signal Strength}
    \acro{GBP}{Gain-Bandwidth Product}

    \acro{LoS}{Line-of-Sight}
    \acro{nLoS}{non-Line-of-Sight}
    \acro{QoS}{Quality of Service}
    \acro{NB}{Narrowband Communication}
    \acro{WPT}{Wireless Power Transfer}
    \acro{IBPT}{Intra-Body Power Transfer}
    \acro{ICNIRP}{International Commission on Non-Ionizing Radiation Protection} 
    \acro{SAR}{Specific Absorption Rate}

    \acro{RF}{Radio Frequency}
    \acro{SNR}{Signal-to-Noise Ratio}
    \acro{RFID}{Radio Frequency Identification}
    \acro{IoT}{Internet of Things}
    \acro{IoUT}{Internet of Underwater Things}
    \acro{UWN}{Underwater Wireless Network}
    \acro{UWSN}{Underwater Wireless Sensor Node}
    \acro{UAC}{Underwater Acoustic Channel}
    \acro{FSK}{Frequency Shift Keying}
    \acro{OOK}{On-Off Keying}
    \acro{ASK}{Amplitude Shift Keying}
    \acro{UUID}{Universal Unique Identifier}
    \acro{PZT}{Lead Zirconium Titanate}
    \acro{AC}{Alternating Current}
    \acro{NVC}{Negative Voltage Converter}
    \acro{NVCR}{Negative Voltage Converter Rectifier}
    \acro{FWR}{Full-Wave Rectifier}
    \acro{GPIO}{General Purpose Input/Output}
    \acro{PCB}{Printed Circuit Board}
    \acro{AUV}{Autonomous Underwater Vehicle}
    \acro{BLE}{Bluetooth Low Energy}
    \acro{FSR}{Force Sensing Resistor}
    \acro{SiP}{System in Package}
    \acro{SoC}{system-on-chip}
    \acro{SpO2}{Oxygen Saturation}
    \acro{PULP}{Parallel Ultra-Low Power}
    \acro{ADC}{analog-to-digital converter}
    \acro{TCDM}{Tightly Coupled Data Memory}
    \acro{GNSS}{Global Navigation Satellite System}
    \acro{IC}{Integrated Circuit}
    \acro{POM}{Polyoxymethylene}
    \acro{RTOS}{Real-Time Operating System}
    \acro{LoRaWAN}{Long Range Wide Area Network}
    \acro{ML}{Machine Learning}
    \acro{IIR}{Infinite Impulse Response}
    \acro{PLL}{Phase-locked Loop}
    \acro{RMS}{Root Mean Square}
    \acro{FPC}{Flexible Printed Circuit}
    \acro{I2C}{Inter Integrated Circuit}
    \acro{UART}{Universal Asynchronous Receiver-Transmitter}
    \acro{DMA}{direct memory access}

    \acro{ToF}{Time of Flight}
    \acro{US}{ultrasound}
    \acro{STR}{Static Target Rejection}
    \acro{RRR}{Reflection Rejection Range}

    \acro{FIR}{Finite Impulse Response}
    \acro{LIDAR}{Light Detection and Ranging}
    \acro{AoA}{Angle of Arrival}
    \acro{FoV}{Field of View}
    \acro{MAE}{Mean Absolute Error}
    \acro{IMU}{Inertial Measurement Unit}

    \acro{CFD}{Computational Fluid Dynamics}
    \acro{FIS}{Fédération Internationale de Ski}
    \acro{GRF}{Ground Reaction Forces}

    \acro{ECA}{Efficient Channel Attention}

    \acro{HV}{high-voltage}
    \acro{LV}{Low Voltage}
    \acro{TX}{transmit}
    \acro{RX}{Receive}
    \acro{TR}[T/R]{Transmit/Receive}
    \acro{AFE}{Analog Front End}
    \acro{LNA}{Low-Noise Amplifier}
    \acro{TGC}{Time-Gain Compensation}
    \acro{PRF}{Pulse Repetition Frequency}
    \acro{ASIC}{Application-Specific Integrated Circuit}
    \acro{CMOS}{Complementary Metal-Oxide-Semiconductor}
    \acro{BCD}{Bipolar-CMOS-DMOS}
    \acro{LDMOS}{Laterally Diffused Metal-Oxide-Semiconductor}
    \acro{ESD}{Electrostatic Discharge}
    \acro{SPI}{Serial Peripheral Interface}
    \acro{FPGA}{field-programmable gate array}
    \acro{DUT}{Device Under Test}
    \acro{FoM}{Figure of Merit}
    \acro{PMUT}{Piezoelectric Micromachined Ultrasonic Transducer}
    \acro{CMUT}{Capacitive Micromachined Ultrasonic Transducer}
    \acro{PVDF}{Polyvinylidene Fluoride}
    \acro{MI}{Mechanical Index}
    \acro{TI}{Thermal Index}

    \acro{IP}{intellectual property}
    \acro{OBI}{Open Bus Interface}
    \acro{RTL}{register-transfer level}
    \acro{PDK}{process design kit}
    \acro{GE}{Gate Equivalent}
    \acro{kGE}{kilo gate equivalents}

    \acrodefplural{SoC}[SoCs]{systems-on-chip}

\end{acronym}

\DeclareSIUnit{\kGE}{kGE}

\newcommand{\ProcNode}{\qty{130}{\nano\meter}}             
\newcommand{\ProcName}{IHP}                                
\newcommand{\CellLib}{ETH EZ130 8T}                        
\newcommand{\ClkFreq}{\qty{100}{\mega\hertz}}              
\newcommand{\SoCName}{Croc}                                
\newcommand{\SoCArea}{\qty{104}{\kGE}}                     

\providecommand{\PulserBase}{\texttt{0x0300\_C000}}         
\providecommand{\CoreStride}{\texttt{0x20}}                 
\providecommand{\PadDrive}{\qty{16}{\milli\ampere}}         
\providecommand{\EndMax}{\num{65535}}                       
\providecommand{\CntMax}{255}                               

\newcommand{\NChanMax}{16}                                 
\newcommand{\NChanInst}{8}                                 
\newcommand{\NChanMCU}{2}                                  

\newcommand{\AreaBase}{\qty{1.65}{\kGE}}                    
\newcommand{\AreaPerChan}{\qty{1.66}{\kGE}}                 
\newcommand{\AreaEightCh}{\qty{14.9}{\kGE}}                  
\newcommand{\AreaFrac}{\qty{14.3}{\percent}}                

\newcommand{\NCycles}{\num{4379}}                          
\newcommand{\FPGABoard}{Terasic DE10-Lite}                 
\newcommand{\Skew}{\qty{10}{\nano\second}}                 

\begin{document}
\bstctlcite{IEEEexample:BSTcontrol} 

\title{U-Sonic: An Open-Source 8-Channel Ultrasound Transmit IP in a 130\,nm RISC-V SoC}

\author{
    \begin{minipage}{0.96\textwidth}\centering
    Federico~Villani\;\IEEEmembership{(Graduate Student Member, IEEE)},
    Nico~Canzani,
    Marc-Andr\'{e}~Wessner \IEEEmembership{(Student Member, IEEE)},
    Philippe~Sauter\;\IEEEmembership{(Graduate Student Member, IEEE)},
    Enrico~Zelioli\;\IEEEmembership{(Graduate Student Member, IEEE)},
    Andrea~Cossettini\;\IEEEmembership{(Senior Member, IEEE)},
    Christoph~Leitner$^\dagger$\;\orcidlink{0000-0002-7058-7236}\IEEEmembership{(Senior Member, IEEE)},
    Luca~Benini\;\orcidlink{0000-0001-8068-3806}\IEEEmembership{(Fellow, IEEE)}
    \end{minipage}
    \thanks{This work was supported in part by the Swiss National Science Foundation (SNSF) under Ambizione Grant 233457.}
    \thanks{F. Villani, N. Canzani, M.-A. Wessner, P. Sauter, E. Zelioli, A. Cossettini, C. Leitner, and L. Benini are with the Integrated Systems Laboratory, ETH Zurich, Switzerland.}
    \thanks{L. Benini is also with the Department of Electrical, Electronic, and Information Engineering, University of Bologna, 40126 Bologna, Italy.}
    \thanks{$^\dagger$~Corresponding E-mail: christoph.leitner@iis.ee.ethz.ch.}
}

\maketitle

\begin{abstract}
Miniaturized \ac{US} probes require programmable and synchronized \ac{TX} excitation across multiple elements, while existing compact platforms often rely on limited \ac{MCU} pulse generators or closed-source fixed-function pulser devices. We present U-Sonic, an open-source digital \ac{US} \ac{TX} peripheral integrated into a 32-bit RISC-V \ac{SoC}. The implemented \ac{SoC} integrates \NChanInst{} pulser cores, while the parameterized architecture supports up to 16 channels. Each core generates single- or dual-tone bursts with programmable period, duty cycle, pulse count, polarity, and idle level, together with optional inverted stop pulses for active damping. A shared memory-mapped \ac{OBI} enables synchronous start and stop of arbitrary channel subsets and supports composite bipolar, gated, and three-level excitation schemes. Functional correctness was verified in Verilator against a Python golden model over \NCycles{} checked cycles across directed and randomized configurations, and confirmed on a \FPGABoard{} \ac{FPGA}. The design was synthesized and placed-and-routed in \ProcName{} \ProcNode{}. The post-layout area in \ac{kGE}, scales as \AreaBase{} plus \AreaPerChan{} per channel. The \NChanInst-channel instance occupies \AreaEightCh{}, corresponding to approximately \AreaFrac{} of the \SoCArea{} \ac{SoC}. The \ac{RTL}, register descriptions, verification collateral, and software support are released as open source.
\end{abstract}


\section{Introduction}\label{sec:intro}

\IEEEPARstart{W}{earable} and miniaturized \ac{US} probes move the \ac{TX} electronics onto the body, where the number of elements and the precision of their excitation determine the capabilities of the front end~\cite{ref_rbme2026}. Coherent firing, aperture selection, and beamforming all require synchronized excitation of several transducer elements~\cite{hu2023cardiac}.

Existing compact US platforms implement \ac{TX} timing through different levels of integration. MCU-based probes typically provide only a small number of pulse-generation channels and often multiplex a single driver across several elements~\cite{ref_ti_msp430fr5043,ref_ius2022,wulpuspro2026,jonveaux2017}. Research ``open platform'' scanners built on \acp{FPGA} generate arbitrary per-channel waveforms using \ac{TX} sequencers implemented in FPGA fabric, but generally do not expose reusable RTL~\cite{boni2018open}. Multi-element discrete front ends delegate pattern and delay generation to fixed-function pulser integrated circuits behind a proprietary register interface and a fixed analog stage~\cite{ti_tx7332,guanziroli2017}. 

\begin{table*}[!t]
    \centering
    \begin{threeparttable}
    \caption{Comparison of ultrasound transmit timing generators.}
    \label{tab:soa_pulsers}
    \footnotesize
    \setlength{\tabcolsep}{3pt}
    \renewcommand{\arraystretch}{1.25}
    \begin{tabular}{@{}>{\raggedright\arraybackslash}m{0.145\textwidth}
                      *{5}{>{\centering\arraybackslash}m{0.125\textwidth}}
                      >{\centering\arraybackslash}m{0.152\textwidth}@{}}
        \toprule

            & \textsc{mcu} class                    
            & Logic only                            
            & Pulser \textsc{ic}                    
            & Digital \textsc{asic}                 
            & \textsc{hv} \textsc{asic}             
            & \textbf{This work}                    
            \\
            & MSP430 \textsc{uss}~\cite{ref_ti_msp430fr5043}
            & LM96570~\cite{ti_lm96570}
            & STHV1600~\cite{st_sthv1600}           
            & Dusa \textit{et al.}~\cite{dusa2015}
            & Tan \textit{et al.}~\cite{tan2020}
            & \textbf{U-Sonic}
            \\

        \midrule

        Realization
            & commercial                                
            & commercial                                
            & commercial                                
            & UMC \qty{130}{\nano\meter}                
            & TSMC \qty{180}{\nano\meter} BCD           
            & \textbf{\ProcName{} \ProcNode}            
            \\

        \acs{TX} channels
            & \NChanMCU, multiplexed                
            & 8                                     
            & \NChanMax                             
            & \NChanMax                             
            & 64                                    
            & \textbf{\NChanInst{} (\NChanMax{} max.)}   
            \\

        Per-channel delay
            & none                                  
            & \qty{0.78}{\nano\second}              
            & \qty{5}{\nano\second}                 
            & \qty{1.25}{\nano\second}              
            & \qty{5}{\nano\second}                 
            & \textbf{none}                         
            \\

        Burst programmability
            & multi-tone, 0--127 pulses             
            & $\le$\,64 pulses                      
            & 64\,kbit pattern memory               
            & length, pattern, frequency            
            & 1--63 cycles                          
            & \textbf{dual-tone, 0--255 pulses, duty, polarity} 
            \\

        Damping stop pulses
            & 0--15, inverted                       
            & n/r                                   
            & return to zero                        
            & n/r                                   
            & return to zero                        
            & \textbf{0--255, inverted}             
            \\

        Drive stage
            & logic                                 
            & logic                                 
            & \qty{\pm100}{\volt}, 5 level          
            & logic                                 
            & \qty{\pm30}{\volt} bipolar            
            & \textbf{logic}                        
            \\

        Programming
            & \textsc{mcu} registers                
            & serial                                
            & serial                                
            & serial (\textsc{ddr})                 
            & shift register, \textsc{lvds}         
            & \textbf{memory-mapped \acs{OBI}}      
            \\

        Design files released
            & $\pmb{\times}$                        
            & $\pmb{\times}$                        
            & $\pmb{\times}$                        
            & $\pmb{\times}$                        
            & $\pmb{\times}$                        
            & $\pmb{\checkmark}$                    
            \\

        \bottomrule
    \end{tabular}

    \begin{tablenotes}[flushleft]
        \footnotesize
        \item \textsc{hv}: high-voltage. n/r: not reported. $\pmb{\checkmark}$ released, $\pmb{\times}$ not released.
    \end{tablenotes}
    \end{threeparttable}
    \vspace{-0.5cm}
\end{table*}

Currently, openness stops at the boundary of the timing device (\cref{tab:soa_pulsers}). Even platforms that release complete board and firmware information still take their excitation timing from silicon that they cannot modify~\cite{ref_ius2022,tinyprobe2025,leitner2025modulus}. Some custom transmitter implementations exist.
Dusa et al.~\cite{dusa2015} reported a 16-channel digital transmit and receive beamformer in a \qty{130}{\nano\meter} process. However, they did not release the design files or a processor-visible register interface. 

In this context, this paper presents U-Sonic, an open-source peripheral that exposes multi-element burst generation as memory-mapped registers of a RISC-V SoC~\cite{croc2025}. The contributions are:
\begin{itemize}[leftmargin=*,topsep=2pt,itemsep=1pt]
  \item A synthesizable \ac{TX} peripheral with up to \NChanMax{} independently programmable, synchronized pulser cores behind an \ac{OBI}. 
  \item A verification flow built on Verilator and a Python golden model, covering directed and random configurations, with functional confirmation on a \FPGABoard{} \ac{FPGA}.
  \item A physical implementation in \ProcName{} \ProcNode{} through an open synthesis and place-and-route flow.

\end{itemize}
The \ac{RTL}, register descriptions, verification documentation, driver software, and physical implementation files are released with a permissive license at {\url{https://github.com/ModularUS/U-Sonic}}.

\section{U-Sonic Transmit Peripheral}\label{sec:design}

\begin{figure}[t]
    \centering
    \newcommand{\FigOneH}{33mm}
    \begin{tabular}{@{}c@{\hspace{3mm}}c@{}}
        \includegraphics[height=\FigOneH]{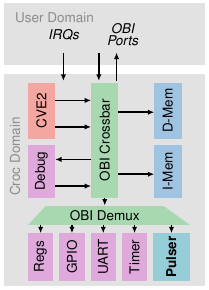} &
        \includegraphics[height=\FigOneH]{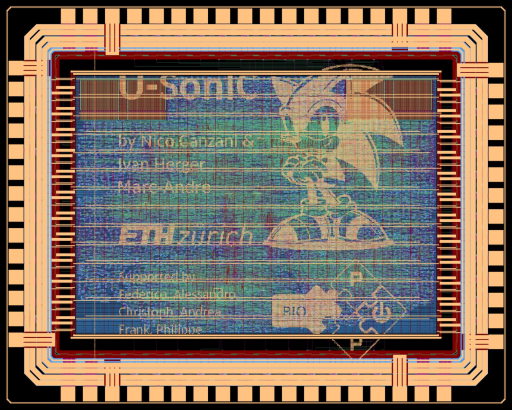} \\[1pt]
        {\footnotesize (a)} & {\footnotesize (b)}
    \end{tabular}\\[3pt]
    \includegraphics[width=\columnwidth]{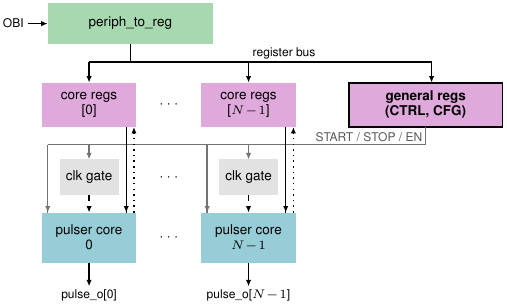}\\[1pt]
    {\footnotesize (c)}
    \vspace{-0.2cm}
    \caption{(a) \SoCName{} \ac{SoC} with the pulser peripheral on the \ac{OBI} demultiplexer, (b) layout in \ProcName{} \ProcNode, (c) pulser architecture; solid: register bus and configuration, gray: START, STOP, and EN strobes, dashed: gated clock, dotted: STATUS write-back.}
    \label{fig:block}
    \vspace{-0.2cm}
\end{figure}

\subsection{System Overview}

Fig.~\ref{fig:block} provides an overview of the U-Sonic \ac{TX} subsystem and its integration into the Croc SoC~\cite{croc2025}. Starting from the \ac{TX} path, U-Sonic comprises \NChanInst{} parallel pulser cores that generate programmable logic-level excitation signals at the chip I/Os. Each core provides an independent burst configuration, while a shared control interface allows any subset of channels to be started or stopped synchronously. The pulser cores are exposed to software through an OBI-to-register bridge, enabling the RISC-V processor to configure the \ac{TX} parameters and trigger multi-channel transmissions through standard memory-mapped accesses.

U-Sonic is integrated as an \ac{OBI} peripheral within the Croc \ac{SoC}, reusing its processor, memory, debug, and software infrastructure.

\subsection{Croc Platform Overview}

Croc is an open-source, 32-bit \ac{MCU}-class \ac{SoC} designed as a compact and extensible platform for education and domain-specific hardware integration~\cite{croc2025}. It is built around the CVE2 RISC-V core from the OpenHW Group, supporting the RV32\{I,E\}[M]C instruction-set configurations and the optional CORE-V eXtension Interface (CV-X-IF) for tightly coupled application-specific extensions~\cite{openhw2022cve2,openhw_cv_xif}.
As shown in Fig.~\ref{fig:block}(a), Croc is partitioned into a \emph{Croc domain}, implementing the reusable host system, and a \emph{user domain} for application-specific extensions. The Croc domain integrates the processor, configurable on-chip SRAM, an optional \ac{DMA} engine, debug support, and a minimal set of memory-mapped peripherals, including UART, GPIO, timers, and system-control registers. These components are connected through a single-cycle \ac{OBI} crossbar, with independent instruction and data interfaces enabling concurrent accesses to separate SRAM banks.

The user domain exposes \ac{OBI} manager and subordinate interfaces, external I/Os, and interrupt lines for integrating custom peripherals, accelerators, and processor extensions. Memory-mapped IPs can be connected as \ac{OBI} subordinates and assigned control and status registers in the processor address space. Analog or mixed-signal blocks, such as an \ac{ADC}, can similarly be integrated through a lightweight memory-mapped wrapper. For higher-throughput peripherals, an \ac{OBI} manager interface can provide direct access to the on-chip SRAM, enabling data transfers without continuous processor intervention. Custom IP can alternatively be instantiated directly in the Croc domain when tighter integration or a fixed system configuration is required.
Croc further provides the software and implementation infrastructure required to develop and validate complete systems. Its bare-metal environment includes startup code, linker scripts, peripheral drivers, interrupt handling, and \ac{DMA} support. Verilator supports RTL simulation and verification, while an FPGA flow targeting the Digilent Genesys~2 enables rapid system-level prototyping. For ASIC implementation, Croc provides an open-source flow based on Bender, Yosys~\cite{wolf2013yosys}, OpenROAD~\cite{ajayi2019toward}, and KLayout. The flow targets IHP's \ProcNode{} \ac{PDK} and is distributed through the IIC-OSIC-TOOLS environment~\cite{croc2025,iic-jku2024iic-osic-tools}. The platform has also been validated in silicon through multiple demonstrators~\cite{zelioli2026croc}.

\subsection{Peripheral and Register Interface}
\label{ssec:system}

U-Sonic is connected as an \ac{OBI} subordinate behind the peripheral demultiplexer of the Croc domain, next to the UART, GPIO, and timer blocks (Fig.~\ref{fig:block}(a)). In the taped-out \ac{SoC}, it occupies a \qty{4}{\kilo\byte} address window starting at \PulserBase{} and drives \NChanInst{} outputs through \PadDrive{} I/O pads.

As shown in Fig.~\ref{fig:block}(c), the peripheral consists of an \ac{OBI}-to-register bridge (\texttt{periph\_to\_reg}), $N$ pulser cores, one register bank and one clock gate per core, and a shared general register bank. Each core occupies a \CoreStride{} register window, while the general register bank follows the $N$ core windows. The shared \texttt{START}, \texttt{STOP}, and \texttt{EN} fields are \NChanMax{} bits wide, defining the maximum number of supported pulser cores. The clock gate associated with each core is controlled by its \texttt{EN} bit, allowing inactive cores to remain static while the register interface continues to operate on the free-running peripheral clock.

\begin{table}[t]
    \caption{Register map. Core registers repeat every \CoreStride{} per core and the general bank sits at offset $N\cdot$\CoreStride{}.}
    \label{tab:regs}
    \centering\scriptsize
    \setlength{\tabcolsep}{3pt}
    \begin{tabular}{@{}lll>{\raggedright\arraybackslash}p{17mm}@{}}
        \toprule
        Offset & Register & Fields [bits] & Function \\
        \midrule
        \multicolumn{4}{@{}l}{\textit{per core}} \\
        0x00 & \texttt{CFG\_F1}   & \texttt{SWITCHVAL} [15:0], \texttt{ENDVAL} [31:16] & F1 high time, period \\
        0x04 & \texttt{CFG\_F2}   & \texttt{SWITCHVAL} [15:0], \texttt{ENDVAL} [31:16] & F2 high time, period \\
        0x08 & \texttt{CFG\_CNT}  & \texttt{F1} [7:0], \texttt{F2} [15:8], \texttt{CNT\_STOP} [23:16] & pulses per phase \\
        0x0C & \texttt{STATUS}    & \texttt{READY} [0], \texttt{STATE} [3:1] & read only \\
        0x10 & \texttt{CTRL\_OUT} & \texttt{INVERT\_OUT} [0], \texttt{IDLE\_OUT} [1] & polarity, idle level \\
        \midrule
        \multicolumn{4}{@{}l}{\textit{shared}} \\
        0x00 & \texttt{CTRL} & \texttt{START} [15:0], \texttt{STOP} [31:16] & write strobe, one bit per core \\
        0x04 & \texttt{CFG}  & \texttt{EN} [15:0] & clock enable per core \\
        \bottomrule
    \end{tabular}
\end{table}

Both the core-local and shared register banks are described in HJSON and translated into SystemVerilog register logic and C headers using the OpenTitan register-generation tool. A corresponding driver in the Croc bare-metal runtime wraps the headers and exposes them to software. 

\begin{figure}[t]
    \centering
    \includegraphics[width=0.86\columnwidth]{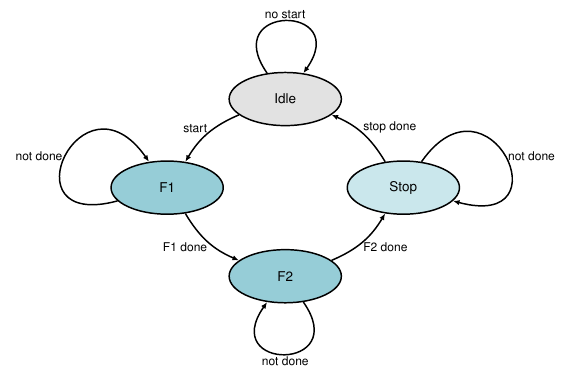}
    \caption{Simplified pulser core state machine. F1 and F2 are the two tone phases of a burst, and Stop is the inverted damping phase. Self-loops hold the state while the current phase has not reached its pulse count. Zero-count phases are skipped, a stop strobe returns the core to Idle from any state.}
    \label{fig:fsm}
\end{figure}

\subsection{Pulser Core}
\label{ssec:core}

Each pulser core is implemented as a five-state machine comprising Idle, the two tone phases F1 and F2, Stop, and Done, together with a 16-bit phase counter and an 8-bit pulse counter (\cref{fig:fsm}). During the F1 and F2 states, the output remains high for $N_{\mathrm{s}}$ clock cycles (\texttt{SWITCHVAL}) within each period of $N_{\mathrm{p}}$ cycles (\texttt{ENDVAL}). The resulting burst frequency and duty cycle are therefore
\begin{equation}
    f_{\mathrm{pulse}} =
    \frac{f_{\mathrm{clk}}}{N_{\mathrm{p}}},
    \qquad
    D =
    \frac{N_{\mathrm{s}}}{N_{\mathrm{p}}},
    \label{eq:burst}
\end{equation}
with a resolution of one clock cycle (\Skew{} at \ClkFreq). $N_{\mathrm{p}}$ spans 2 to \EndMax{} cycles, while each phase can generate between 0 and \CntMax{} pulses. Setting the pulse count of either phase to zero skips that phase, allowing both single-tone and dual-tone bursts to be generated.

After the active burst phases, \texttt{CNT\_STOP} can append inverted periods of the last active phase to reduce transducer ringdown. \texttt{INVERT\_OUT} controls the polarity of the active phases, while \texttt{IDLE\_OUT} defines the output level in the Idle and Done states. The \texttt{STATUS} register exposes the current state and a \texttt{READY} flag, which is polled by software.

Configuration registers are read directly by the core and must therefore remain unchanged during an active burst. Start and stop commands are generated from the shared control register. A write to \texttt{CTRL} asserts the corresponding control signal for all selected cores in the same cycle, with each selected core leaving Idle on the following clock edge. The stop command has priority over the active states and returns the core to Idle within one cycle while resetting both counters.

\subsection{Multi-Channel Synchronization and Drive Modes}
\label{ssec:modes}

Since a single write to the shared \texttt{CTRL} register controls all pulser cores, any subset of the $N$ outputs can be started or stopped on the same clock edge. This avoids sequential software triggering and provides cycle-accurate synchronization across the selected channels.

\begin{figure}[t]
    \centering
    \includegraphics[width=0.8\columnwidth]{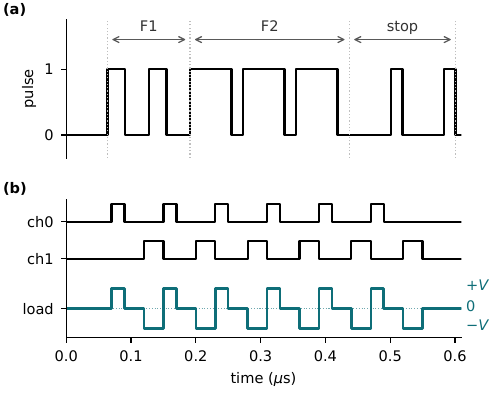}
    \vspace{-0.3cm}
    \caption{(a) Burst phases of one pulser core, F1, F2, and inverted stop pulses. (b) Two cores started on the same clock edge produce a three-level drive across the load, using different switch points and the invert flag rather than any per-channel delay.}
    \label{fig:waveform}
        \vspace{-0.3cm}
\end{figure}

Combining independently configured cores with synchronous triggering enables several drive schemes, as illustrated in \cref{fig:waveform}. Examples include: (i) a bipolar burst from two cores with opposite \texttt{INVERT\_OUT}; (ii) a gated burst from one core with a long gate (\texttt{SWITCHVAL} close to \texttt{ENDVAL}) while a second core bursts inside it; and (iii) a three-level drive across a load connected between two outputs, using two cores with different \texttt{SWITCHVAL} settings and an inverted output, yielding $+V$, $0$, and $-V$.

The current peripheral generates periodic dual-tone bursts on a common time base. It has no per-channel delay field, no pattern memory or drive stage. Focusing and steering require the addition of a delay stage and are left as a future extension.

\section{Implementation and Results}\label{sec:results}

Functional correctness was verified in Verilator against a Python golden model using directed and randomized configurations. Across \NCycles{} checked cycles, the RTL output matched the reference model without discrepancies. The peripheral was additionally implemented on a \FPGABoard{} \ac{FPGA}, where a representative subset of pulse configurations was generated and the resulting waveforms and multi-channel synchronization were observed in hardware.

The design was synthesized and placed-and-routed in \ProcName{} \ProcNode{} with the \CellLib{} standard-cell library~\cite{iip2026ezlibrary}, using Yosys and OpenROAD. Gate equivalents refer to the NAND2 cell of that library. Post-layout area scales approximately linearly with the number of instantiated channels as
\begin{equation}
    A_{\mathrm{pulser}}(N) \approx \AreaBase{} + N \cdot \AreaPerChan{}.
    \label{eq:area}
\end{equation}
The \NChanInst-channel instance occupies \AreaEightCh{}, corresponding to approximately \AreaFrac{} of the \SoCArea{} RISC-V \ac{SoC}. While the implemented instance contains \NChanInst{} channels, the peripheral architecture supports up to \NChanMax{} synchronized \ac{TX} channels.

\Cref{tab:soa_pulsers} compares U-Sonic with representative \ac{US} \ac{TX} timing generators across different integration levels. U-Sonic provides programmable single- and dual-tone bursts, duty cycle and polarity control, inverted stop pulses, and synchronous multi-channel triggering through a memory-mapped \ac{OBI} interface. Unlike the compared implementations, its \ac{RTL} and register descriptions are openly released, enabling modification and reuse within Croc-based systems. The current implementation does not include per-channel delay control or a high-voltage drive stage.

\section{Conclusion}\label{sec:conclusion}
This work demonstrates a compact and reproducible \ac{TX} subsystem for multi-element \ac{US} front-end research. The design was functionally validated through RTL verification and \ac{FPGA} prototyping, and it was physically implemented in \ProcName{} \ProcNode{}. Future work will focus on silicon characterization of the taped-out \ac{SoC} and the integration of a programmable per-channel delay stage for \ac{TX} focusing and steering. The complete \ac{RTL}, register descriptions, verification collateral, and software support are released as open source at {\url{https://github.com/ModularUS/U-Sonic}}.

\section*{Acknowledgment}
{\small We thank Ivan Herger for his contribution to the chip design, Alessandro Ottaviano for supervising the pulser development, and the Microelectronics Design Center at ETH Zurich for support with chip fabrication.}

\bibliographystyle{IEEEtran}
\bibliography{bib/references}

\end{document}